\documentclass[twocolumn,prl,aps,floats,amsfonts,amssymb,showpacs,superscriptaddress]{revtex4}
\usepackage[utf8]{inputenc}
\usepackage{graphicx}
\usepackage{color}
\usepackage{psfrag}
\usepackage{gensymb}
\usepackage{enumerate}
\usepackage{epstopdf}
\usepackage{amssymb}
\usepackage{amsmath}
\usepackage{wasysym}
\usepackage{hyperref}
\usepackage{color}

\newcommand{\symb}[2]{$($\textcolor{#1}{#2}$)$}
\newcommand{\sq}{\tiny $\blacksquare$}
\newcommand{\ltri}{$\blacktriangleleft$}
\newcommand{\rtri}{$\blacktriangleright$}
\newcommand{\utri}{$\blacktriangle$}
\newcommand{\dtri}{$\blacktriangledown$}
\newcommand{\bstar}{\scriptsize $\bigstar$}
\newcommand{\diam}{\scriptsize $\blacklozenge$}
\newcommand{\plus}{$(+)$}		% Ca ̩vite les d̩formations du symbole dans les l̩gendes
\newcommand{\minus}{$(-)$}		% Permet une longueur plus "normale" du signe -
\begin{document}

%\title{Dissipative anomaly in a fluid and superfluid von Karman Helium experiment}
\title{Probing quantum and classical turbulence analogy through global bifurcations in a von K\'arm\'an liquid Helium experiment}

\author{SHREK Collaboration: B. Saint-Michel}
\affiliation{Laboratoire SPHYNX, CEA/IRAMIS/SPEC, CNRS URA 2464, F-91191 Gif-sur-Yvette, France}

\author{E. Herbert}
\affiliation{Laboratoire FAST, CNRS UMR 7608, Universit\'e Paris-Sud, Universit\'e Pierre-et-Marie-Curie, B\^at. 502, Campus universitaire, 91405 Orsay, France}

\author{ J. Salort}
\affiliation{Laboratoire de Physique de l'\'ENS de Lyon, CNRS/Universit\'e Lyon - F-69364 Lyon cedex 7, France}

\author{C. Baudet} 
\affiliation{Laboratoire des \'Ecoulements G\'eophysiques et Industriels, CNRS/UJF/INPG, F-38041 Grenoble Cedex 9, France}

\author{M. Bon Mardion}
\affiliation{Service des Basses Temp\'eratures, INAC/SBT, UMR CEA-UJF 9004, CEA Grenoble, 17 rue des Martyrs 38054 Grenoble Cedex France}

\author{P. Bonnay}
\affiliation{Service des Basses Temp\'eratures, INAC/SBT, UMR CEA-UJF 9004, CEA Grenoble, 17 rue des Martyrs 38054 Grenoble Cedex France}

\author{M. Bourgoin}
\affiliation{Laboratoire des \'Ecoulements G\'eophysiques et Industriels, CNRS/UJF/INPG, F-38041 Grenoble Cedex 9, France} 

\author{B. Castaing}
\affiliation{Laboratoire de Physique de l'\'ENS de Lyon, CNRS/Universit\'e Lyon - F-69364 Lyon cedex 7, France}

\author{L. Chevillard}
\affiliation{Laboratoire de Physique de l'\'ENS de Lyon, CNRS/Universit\'e Lyon - F-69364 Lyon cedex 7, France}

\author{F. Daviaud}
\affiliation{Laboratoire SPHYNX, CEA/IRAMIS/SPEC, CNRS URA 2464, F-91191 Gif-sur-Yvette, France}

\author{P. Diribarne}
\affiliation{Service des Basses Temp\'eratures, INAC/SBT, UMR CEA-UJF 9004, CEA Grenoble, 17 rue des Martyrs 38054 Grenoble Cedex France}

\author{B. Dubrulle}
\affiliation{Laboratoire SPHYNX, CEA/IRAMIS/SPEC, CNRS URA 2464, F-91191 Gif-sur-Yvette, France}

\author{Y. Gagne}
\affiliation{Laboratoire des \'Ecoulements G\'eophysiques et Industriels, CNRS/UJF/INPG, F-38041 Grenoble Cedex 9, France} 

\author{M. Gibert}
\affiliation{Universit\'e Grenoble Alpes, Institut N\'EEL, F-38042 Grenoble, France,\\
CNRS, Institut N\'EEL, F-38042 Grenoble, France}

\author{A. Girard}
\affiliation{Service des Basses Temp\'eratures, INAC/SBT, UMR CEA-UJF 9004, CEA Grenoble, 17 rue des Martyrs 38054 Grenoble Cedex France}

\author{B. H\'ebral}
\affiliation{Universit\'e Grenoble Alpes, Institut N\'EEL, F-38042 Grenoble, France,\\
CNRS, Institut N\'EEL, F-38042 Grenoble, France}

\author{Th. Lehner}
\affiliation{LUTH, Observatoire Paris-Meudon, 5 Pl. Jules Janssen, F-92195 Meudon Cedex, France}

\author{B. Rousset}
\affiliation{Service des Basses Temp\'eratures, INAC/SBT, UMR CEA-UJF 9004, CEA Grenoble, 17 rue des Martyrs 38054 Grenoble Cedex France}


\begin{abstract}
We report measurements of the dissipation in the Superfluid Helium high REynold number von Karman flow (SHREK) experiment for different forcing conditions,  through a regime of global hysteretic bifurcation. 
Our macroscopical measurements indicate no noticeable difference between the classical fluid and the superfluid regimes, thereby providing evidence of the same dissipative anomaly  and response to asymmetry in fluid and superfluid regime.  %In the latter case, 
A detailed study of the variations of the hysteretic cycle with Reynolds number supports the idea that (i) the stability of the bifurcated states of classical turbulence in this closed flow is partly governed by the dissipative scales and
(ii) the normal and the superfluid component at these temperatures (1.6K) are locked down to the dissipative length scale. 

%at large scales can be described by classical formulae describing conventional turbulence provided the density and the viscosity are chosen respectively as the total helium density and the viscosity of the normal fluid. 
\end{abstract}

\maketitle

%\section{Introduction}

At low temperatures ($T\le 2.17$~K at saturated vapor pressure), liquid Helium 4 is subject to a phase transition, from a classical fluid phase HeI ($T> 2.17$~K at saturated vapor pressure) to a superfluid one HeII which behaves like a two-fluid system, with one normal component, following classical Navier-Stokes equations, and one superfluid irrotational component, with quantized
small scale vortices  and zero viscosity (see e.g. \cite{SkrSre12}). Both components interact with these vortices, which gives raise to the mutual friction between these components.

% 
% This two-fluid behavior explains why Helium 4 ``leaks" through thin capillaries as if it had zero viscosity, or the existence of two modes of longitudinal wave propagations (two sounds). 

When a superfluid Helium 4 flow is forced within a tank or past an obstacle, it can become turbulent. An important open question is how much analogy such ``quantum turbulence" bears with classical turbulence, and how much it can be explained within the above mentioned two-fluid model, and its mutual friction term \cite{skrbek2011}. 

In classical turbulence, at large enough Reynolds numbers, one distinguishes the (large) ``inertial'' scales, where viscosity has no influence, and the (small) ``dissipative'' scales, where viscosity matters. For instance, the energy dissipation is known to be independent of the Reynolds number, thus of the viscosity. It is controlled by the way the large scales feed the intermediate inertial scales. Also, the energy spectrum of isotropic homogeneous turbulence has an universal behavior as $k^{-5/3}$ when the wave number $k$ belongs to the inertial scales.

Since the pioneering von K\'arm\'an liquid Helium experiment of Ref. \cite{tabeling}, it is known that the inertial scales energy spectrum does not change when the fraction of normal to superfluid is lowered from $100$\% to $20$\% and follows the above classical $k^{-5/3}$-law. This finding was later  confirmed in different geometries and with different forcing in~\cite{salort} using both a superfluid jet and grid turbulence. Also, this analogy between the normal and quantum turbulence extends to the energy dissipation, that was found equal in both the classical fluid HeI and the superfluid regime HeII at large Reynolds numbers~\cite{Abid1998,Smith1999,Blazkova2007,Bradley2009,Bradley2012}. Since then, measurements of the energy transfers in superfluid helium through Pitot tubes~\cite{Sal12} have confirmed that they are independent of the turbulence nature (quantum or classical) down to the smallest experimentally accessible scales, unfortunately much larger than dissipative scales~\cite{DuriPhd}. 

Summarizing, the present experimental state-of-the-art provides strong evidence that quantum turbulence can be described at large scales by classical formulae describing conventional turbulence.
% , provided the density and the viscosity are chosen respectively as the total helium density and the viscosity of the normal fluid \cite{Walstrom1988,Baggaley2011}. 
A traditional explanation is that mutual friction tends to lock the normal and superfluid velocity fields together, so that the whole is indistinguishable from a classical fluid with the total helium density and the viscosity of the normal component~\cite{Roche2009}. However, this locking should stop at small scales, when its viscosity prevents the normal component velocity from following the spatial variations of the superfluid component.
% 
% 
% \textcolor{red}{This raises the new interesting question of whether at lower temperature ($T\le 1K$), when the viscous component is negligible, the quasi-classical behavior of the flow prevails. Encouraging experimental efforts in this direction \cite{BradleyNature2011} seem the support this picture.}

In the present letter, we report an even more challenging test of the analogy between quantum and classical turbulence by exploring the response of the system to a forcing asymmetry in the SHREK experiment, a low-temperature {\sl forced} von K\'arm\'an liquid Helium experiment providing very high Reynolds measurements in liquid helium both in its classical (above $2.2$~K) and superfluid phase (below $2.1$~K). This response has been previously extensively studied in a scale 1:4 version of the experiment involving classical fluids (water or glycerol)~\cite{cortet11,ravelet2004,brice} and was shown to lead to global bifurcation akin to first or second order phase transition. In the latter case, the transition involves both inertial effects and non-trivial Reynolds dependence~\cite{brice}, that makes it suitable to explore possible influence of the nature of the dissipation mechanism, and provides more stringent tests on the two-fluid model and of the analogy between quantum and classical turbulence.

%\section{Generalities}
\paragraph*{Experimental setup}
Our experimental setup is described in~\cite{rousset}. It consists of a cylinder of radius $R = 0.39$~m and height $H = 1.2$~m. The fluid is mechanically stirred by a pair of coaxial impellers rotating in opposite direction. The impellers are disks of radius $0.925R$, fitted with 8 radial blades of height $0.2R$ and curvature radius $0.4625R$. Two senses of rotation can be defined, according to whether the flow is pushed with the convex \plus{} or concave \minus{} part of the blades. The disks inner surfaces are $1.8R$ apart setting the axial distance between impellers from blades to blades to $1.4R$. The impellers rotation rate $f_1$ and $f_2$ can be varied independently from $0.1$
to $2$~Hz, delivering a total power ranging from $120$~W at $1.6$~K to $800$~W at $4.5$~K. This experiment benefits from high flexibility of  flow conditions, due to the large variation of helium properties over the available temperature range ($1.6$~K to $5$~K). Both superfluid and normal turbulence measurements are possible in the same experiment, with adjustable fraction of the superfluid component (from $\approx 85\%$ at $1.6$~K to $0\%$ above $2.2$~K). Torque measurements at each impeller are performed with SCAIME technology and provide values over the kHz range of $C_1$ and $C_2$, being respectively the torque applied to the bottom and top shafts. Following the procedure described in~\cite{MariePhD}, they are calibrated using measurements at different mean frequencies, so as to remove spurious contributions from genuine offsets or mechanical frictions.
 
\paragraph*{Control parameters and diagnostics }
The control parameters of the studied von
K\'{a}rm\'{a}n flow are the Reynolds number $Re= \pi (f_1+f_2)
R^2/\nu$, where $\nu$ is the fluid viscosity, which controls the
intensity of turbulence and the rotation number
$\theta=(f_1-f_2)/(f_1+f_2)$, which controls the forcing asymmetry. 
Choosing $R$ and $\Omega^{-1}=(\pi(f_1+f_2))^{-1}$ as unit of length and time, we compute the non-dimensional value $K_{p1}$ and $K_{p2}$ of the torque as: $K_{pi}=C_i/(\rho R^5\Omega^2)$, where $\rho$ is the density of the working fluid. From this, we compute two diagnostics: the mean non-dimensional torque: $ K_p=(Kp_1+Kp_2)/2$, and the response to asymmetry: $\Delta K_p=
(Kp_1-Kp_2)$.  In the exact counter-rotating case ($\theta=0$), the non-dimensionalized mean dissipation then reads $(C_1+C_2)\Omega/(\rho R^5 \Omega^3)=2K_p$. 
In the sequel, we present results obtained using 3 types of fluids, under operating conditions and characteristics that are summarized in Table \ref{tablefluid}. In the superfluid regime, the value of the viscosity is ill-defined, due to the quantum nature of the fluid. In all subsequent comparisons, we adopt the conventional view and assign an {\sl effective} kinematic viscosity $\nu_{\rm eff}=\mu/\rho$, where $\mu$ is the dynamical viscosity of the normal component, and $\rho$ the total density of the fluid, as if both component were locked together. In the table, we also include the characteristics of the scale 1:4 water experiment, that will be used for reference and discussions.

\begin{table}[h]
\begin{center}
\begin{tabular}{|c|c|c|c|c|c|c|c|}
\hline
Fluid & $P$  &  $T$ &$F$ &$R$ & $\rho$ &$10^7 \times \nu$   &Re\\
 & (bar) &   (K) &(Hz) &(m) & (kg$/$m$^3$) &(m$^2/$s)   & \\
\hline
  HeI & 1.1 & 2.62  &0.1-0.6&$0.4$ &147.3 &$0.20$   &$5\,10^{6}$ -- $3\,10^{7}$\\
  HeII & 1.1 & 1.63  &0.1-0.6&$0.4$ & 147.3 &$0.10$  & $1\,10^{7}$ -- $6\,10^{7}$\\
  N$_2$& 3.7 & 284 &1.8&$0.4$ & 4.4 &$40$   &$5\,10^{5}$\\
  H$_2$O & 1.8 & 300 &2-15&$0.1$ & 1000 &$10$   &$1\,10^{5}$ -- $1\,10^{6}$\\

\hline
\end{tabular}
\end{center}
\caption{Summary of the four fluid configurations studied in the present paper, and their main properties
(density $\rho$, viscosity $\nu$, Reynolds number Re) under
operating conditions (Pressure $P$, Temperature $T$, mean impeller frequency $F=(f_1+f_2)/2$ and cylinder radius $R$). In the HeII superfluid regime, the effective viscosity is reported, $\nu_{\rm eff}$.}
\label{tablefluid}
\end{table}
%-----------------------------------------------------------------------------------------------------

%\begin{figure}
%\setlength{\unitlength}{1cm}
%\begin{center} 
%\includegraphics[width=8.9cm]{SHREK-Montage.pdf}
%\end{center}
%\caption{Schematic view of the experimental setup and the impellers blade profile. The arrow on the shaft indicates the \plus{} impeller rotation direction.}
%\label{fig3}
%\end{figure}

A first campaign of experiments took place in October 2012. During this campaign, results were obtained at $P=1.1$ bar above the superfluid transition $T=2.62$~K, and below $T=1.63$~K, to ensure that the total fluid density is equal to $\rho=147.3$~kg/m$^3$ in each case. In the HeII regime, the  superfluid fraction was about 85\%. The low temperature results are compared 
with those obtained in nitrogen at $284$~K and 4 bar. 
The present results therefore span a range of control parameter corresponding to  ${\rm Re} \in[10^5; 10^8]$, and $\theta\in[-1;1]$ for the \minus{} sense of rotation, and 
${\rm Re} \in[10^5; 10^8]$ at $\theta=0$ for the \plus{} sense.

%\section{Results}

\begin{figure}
\setlength{\unitlength}{1cm}
\begin{center} 
\includegraphics[width=8.5cm]{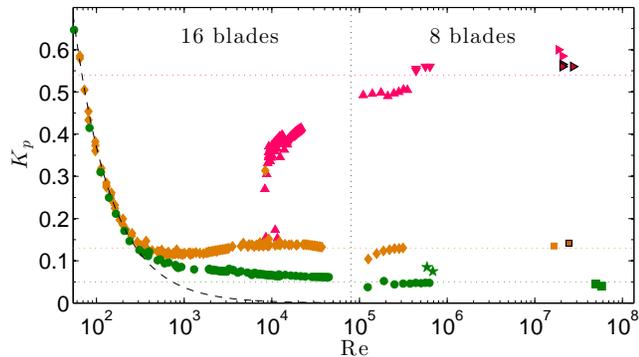}
\end{center}
\caption{Variations of $K_p = (K_{p1} + K_{p2})/2$ as a function of Re at $\theta = 0$ for various fluids. Measurements at the lowest Reynolds numbers are taken from~\cite{ravelet2008} and were obtained using impellers with 16 blades. Measurements at ${\rm Re} \le 10^{5}$ are performed with Polycarbonate (Saclay experiment) or metal (SHREK experiment) impellers with 8 blades, strictly homothetic. Pink \symb{black}{\utri}, bifurcated \minus{} regime in Saclay experiments; pink \symb{black}{\dtri}, same regime in SHREK (nitrogen); pink \symb{black}{\rtri}, experiments in helium. Symbols with black outline account for superfluid experiments. Orange \symb{black}{\diam}, symmetrical \minus{} regime of Saclay experiments. Orange \symb{black}{\sq}, same regime in helium, in SHREK. Green \symb{black}{$\bullet$} and \symb{black}{\bstar}, \plus{} regimes, respectively for Saclay and SHREK nitrogen experiments. Green \symb{black}{\sq}, SHREK helium experiments. Dotted lines are eye-guides. Black dashed line account for viscous stirring, $K_p \propto Re^{-1}$.}
\label{fig4}
\end{figure}

\paragraph*{Dissipation anomaly}
As mentionned above, there is now overwhelming evidence that in classical homogeneous and isotropic turbulent flows of typical fluctuating velocity $u$ and typical length $L$, the mean energy dissipation rate does not vanish in the limit of vanishing viscosity, but instead converges to a finite universal limit when adimensionalized by $u^3/L$~\cite{Fri95,Sreenivasan98} that can be related to the so called Kolmogorov constant~\cite{Pope00,Chevetal}. There is presently no rigorous derivation of this property from the Navier-Stokes equations, so we cannot rule out a priori a dependence of this limit on the nature of energy dissipation. For a non homogeneous and non isotropic flow such as the von K\'arm\'an flow, the adimensionalized mean energy dissipation is given by $2K_p$ and the question is whether $K_p$ undergoes a variation when the superfluid component is becoming more and more important.
Previous measurements of the non-dimensional dissipation $K_p$ in von K\'arm\'an experiments with similar impellers operated at $\theta=0$ for Reynolds numbers between $100$ and $10^6$ are summarized in Fig.~\ref{fig4}, showing 
$K_p$ as a function of Re. They show a saturation of $K_p$ above ${\rm Re}\sim 10^5$ in both rotation senses for counter-rotating impellers at $\theta=0$~\cite{ravelet2008}. The present measurements at $\theta=0$, extending up to ${\rm Re}=6 \times 10^7$, confirm this saturation without any ambiguity, in the three stationary regimes: the symmetric \plus{} and \minus{} and the non-symmetric (``bifurcated") \minus{} regime, see Fig.~\ref{fig4}. The overlap of measurements in N$_2$ and in H$_2$O, performed at similar Reynolds numbers in two different experiments confirms the validity of our calibrations and justifies that no additional bearing due to mechanical frictions can explain this saturation. The dissipation in the superfluid regime (symbols with black outline) for the \minus{} sense is reported in Fig.~\ref{fig4}, using the effective Reynolds number.
One sees that the dissipation is identical in the fluid and in the superfluid regimes, thereby confirming earlier results obtained in a von K\'rm\'an driven by straight-blade impellers~\cite{Abid1998}. 

\begin{figure}
\setlength{\unitlength}{1cm}
\begin{center} 
\includegraphics[width=8.5cm]{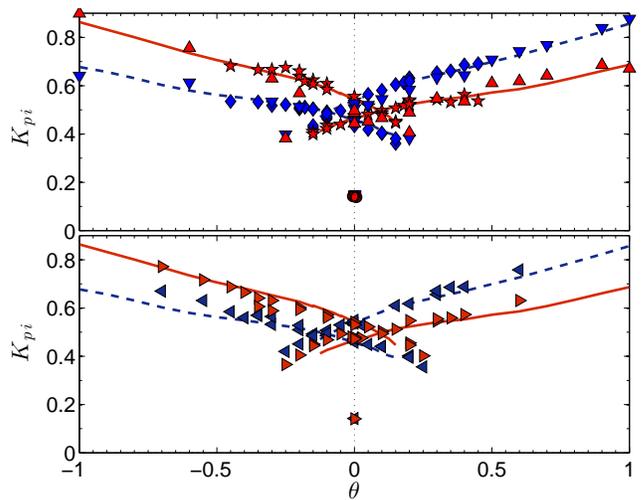}
\end{center}
\caption{Non-dimensional torques $K_{pi}$ corrected with $\theta = 0$ value to ensure $\theta \to -\theta$ symmetry, varying with $\theta$ for various fluids, corresponding to different Reynolds number ranges (see Table I). Left: blue \symb{black}{\diam} and red \symb{black}{\bstar} are respectively the lower and upper torques in nitrogen experiment. Blue \symb{black}{\dtri} and red \symb{black}{\utri}, same quantities in HeI. Right: blue \symb{black}{\rtri} and red \symb{black}{\ltri}, same quantities in HeII. For both graphs, the red solid and the blue dashed lines recall the values of $K_p(\theta)$ for Saclay experiments in water.}
\label{fig5}
\end{figure}

We have further investigated the universality of the energy dissipation as a function of $\theta$ by imparting a differential rotation of the impellers. When $\theta\neq 0$, the torques applied by the two impellers can become different. It is therefore useful to study both $K_{pi}$ individually. This is done in the classical regime in Fig.~\ref{fig5}-Top, for the experiment operating  in the \minus{} sense and for the 3 different fluids, at different Reynolds numbers. All the curves collapse approximately on two master curves describing the torque on the two impellers. For large enough $\theta$, the two curves are single valued, and are ordered with respect to the impeller frequency, the largest non-dimensional dissipation being observed for the impeller rotating at the highest frequency. In the range $\vert\theta\vert\le 0.4$, the dissipation becomes multi-valued, tracing the multistability of large-scale flow solutions. In that interval of $\theta$, the collapse of the curve is not as good as for large $\theta$, since the H$_2$O and N$_2$ fluids appear to be multi-valued only on a smaller interval than Helium I. This difference will be further discussed in the next Section. Overall, the collapse observed in Fig.~\ref{fig5} confirms that the non-dimensional dissipation in the classical regime is universal above ${\rm Re} \ge 10^5$, at any $\theta$. For comparison, we provide in Fig.~\ref{fig5}-Bottom
the torque measurements for various $\theta$ in the regime with 85 \% superfluid. One sees that they collapse on the same two master curves describing the fluid regime.
 In this context, while it is tempting to call for singularities to explain anomalous dissipation in homogeneous and isotropic turbulence, as initially suggested by Onsager \cite{Onsager49,Duchon00,Eyink06}, it appears that the large-scale mean flow is instead of crucial importance to determine the limiting value of $K_p$ in a von K\'arm\'an flow. We can also infer that the very nature of the viscous dissipation does not itself select the large scale flow, justifying a posteriori the statistical physics descriptions of the von K\'arm\'an steady states \cite{monchaux,cortet11}.
 
\begin{figure}
\setlength{\unitlength}{1cm}
\begin{center} 
\includegraphics[width=8.5cm]{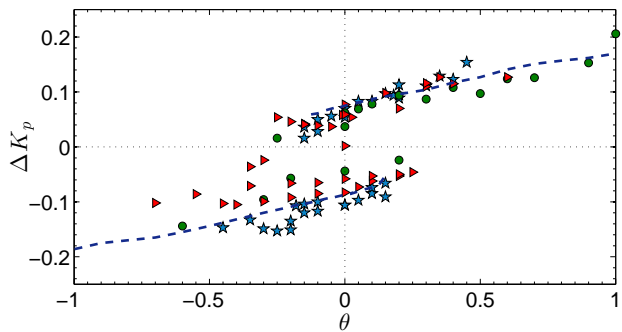}
\end{center}
\caption{Shifted values of $\Delta K_p(\theta)$ in SHREK to ensure symmetry with respect to $\theta = 0$ (same process as in Fig.~\ref{fig5}). Teal \symb{black}{\bstar}, nitrogen experiment~; green \symb{black}{$\bullet$} liquid helium~; red \symb{black}{\rtri}, superfluid helium. Blue dashed line recalls the $\Delta K_p$ values of the Saclay water experiments.}
\label{fig6}
\end{figure}

 \paragraph*{Torque Asymetry response}
 In the \minus{} sense, the torque asymetry actually traces a hysteretic global bifurcation~\cite{ravelet2004,ravelet2008}, corresponding to a spontaneous symmetry breaking between two large scale circulations: (i) a symmetric  flow composed of two symmetric toric
recirculation cells separated by an azimuthal shear layer  located
at $z=0$. This flow is characterized by $K_{p1}=K_{p2}$ and is only observed in the vicinity of $\theta=0$; it is stable at $\theta=0$ and metastable otherwise;  (ii) a  one-cell flow with  a shear layer concentrated in between the blades of the slowest impeller. This flow can coexist with its parity symmetric in a range of parameter $\vert\theta\vert\le \theta_*$. The corresponding torque  $\Delta K_p$ as a function of $\theta$ is made of two parallel branches, the lower (resp. upper) one stopping (resp. starting) at $\theta_*$ (resp. $-\theta_*$), thereby drawing an hysteresis cycle (see dotted line of Fig. \ref{fig6}). 

\begin{figure}
	\centering
	\includegraphics[width = 0.48 \textwidth]{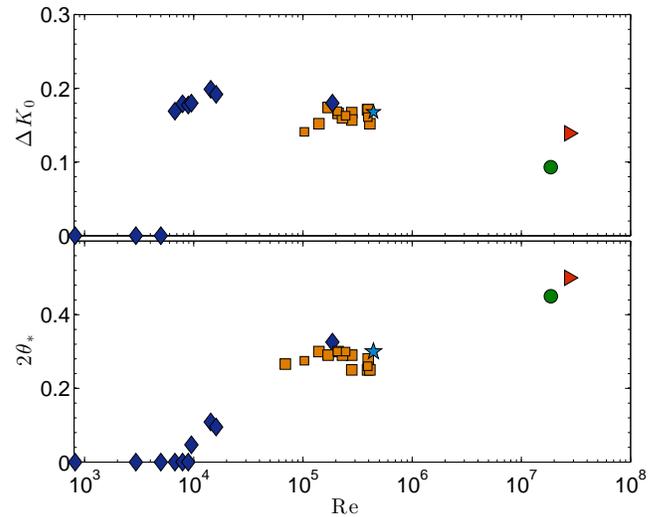}
	\caption{Top: height $\Delta K_0$ of the hysteresis loop plotted in fig.~\ref{fig6}, as a function of the Reynolds number. Bottom: width $2 \theta_*$ of the hysteresis cycle. Blue \symb{black}{\diam} (Orange \symb{black}{\sq}), Saclay experiment in water with 16 blades (8 blades) impellers; teal \symb{black}{\bstar}, SHREK experiment in nitrogen; \symb{black}{$\bullet$}, in liquid helium; red \symb{black}{\rtri}, in superfluid helium.}
	\label{fig7}
\end{figure}

To explore the cycle properties in our SHREK experiment, we have used the data of Fig.~\ref{fig5} to study the torque asymmetry response for different fluids. This is done in Fig.~\ref{fig6}. The data are more scattered than for individual torques, but overall, they collapse around the two straight branches of the hysteresis cycle. The Reynolds dependence of the hysteresis cycle is observed for two quantities: its width,  $2 \theta_*$ and and its height $\Delta K_0$. We have measured these quantities, and plotted them in Fig.~\ref{fig7} that summarizes their behavior for all the fluids. For Reynolds numbers around ${\rm Re } \sim 5\times 10^5$ (H$_2$O and N$_2$), the width extends over roughly $2 \theta_* = 0.3\pm 0.03$ while at ${\rm Re}\sim 2\times 10^7$ (HeI), the width extends up to $2 \theta_*=0.45\pm 0.03$ and the height $\Delta K_0$ is slightly smaller than at lower Reynolds numbers. This trend is confirmed in the superfluid case (HeII), with a hysteresis width also of the order of again $2 \theta_*= 0.5\pm 0.03$ and height of $\Delta K_0= 0.13\pm 0.03$. Note that the precision of the SHREK measurements is not as good as in H$_2$O for $\Delta K_0$, but is good for $2 \theta_*$, which refers to rotation frequencies, measured with a good precision. Overall, one observes a well-defined trend as a function of Re, confirming the previous results obtained in~\cite{brice}. As far as classical fluids are concerned, the present experiment demonstrates that the width of the hysteresis cycle increases with Reynolds numbers leading to the following picture at infinite Reynolds numbers: once bifurcated in one state, the flow remains stuck in it. It also shows that the origin of the transition from one branch to the other when $\theta$ is increased cannot be attributed to the sole inertial scales, as it depends on the viscosity. Moreover, the values observed in the superfluid experiment and plotted using the effective Reynolds number fit well with the classical fluid experiments. This analogy between quantum and classical turbulence is shown here for the first time on a Reynolds dependent quantity. It suggests that the normal and superfluid components are locked together down to the dissipative scales, at least in this temperature range, in agreement with numerical simulations of the two-fluid model \cite{Sal11}.

\paragraph*{Discussion}
Exploring the properties of a global bifurcation in liquid Helium 4 experiment, we have been able to probe the analogy between forced classical and superfluid turbulence in a von K\'arm\'an geometry. Our first result is the confirmation, for a vast range of forcing conditions, that the injected energy in the flow is dissipated in the same way in the purely viscous flow (i.e. $T>2.18$~K) and in the two-components fluid (i.e. $T<2.18$~K) flow. This is expected if the energy transfers are controlled by the large scales where the two components are locked by the mutual friction in the superfluid phase. This also confirms that the very nature of the microscopic dissipation mechanism does not itself select the large scale flow in von K\'arm\'an experiments, as it has been proposed in a statistical physics point of view~\cite{cortet11}. Through a detailed analysis of the global bifurcation properties, we have also been able to detect the signature of an effective dissipative process in the superfluid regime. This signature is comparable to that of a classical Helium 4 with the same total density and the dynamical viscosity of the normal component, that corresponds to $\nu_{\rm eff}\sim 9.4.10^{-9}$~m$^2$/s. 

Such a classical turbulent fluid would have a dissipative scale $\eta_{\rm eff}=(\nu_{\rm eff}^3/\epsilon)^{1/4}$, where $\epsilon$ is the rate of energy dissipation per unit mass. At sufficiently low temperatures, one expects the normal and superfluid components to decouple at the normal component dissipative scale $\eta_n \gg \eta_{\rm eff}$, while the average distance between quantized vortices is $\eta_s \leq \eta_{\rm eff}$~\cite{Sal11}. Depending on the nature of the mechanism which controls our hysteresis, one or the other of these two lengths $\eta_n$ or $\eta_s$ should be pointed out, not $\eta_{\rm eff}$. It is why our results suggest that both components are locked down to the dissipative scale, even if the normal component represents only 15\% of the total density.

{\bf Acknowledgments}\
We thank CNRS and contract ANR-09-BLAN-0094-01 for support, F. Bancel, L. Monteiro, Ph. Charvin, C. Gasquet and V. Padilla for technical support, and Ph.-E. Roche for fruitful discussions.

\end{document}